\documentclass[aip,preprint,onecolumn,showpacs,preprintnumbers,amsmath,amssymb,superscriptaddress,floatfix]{revtex4-2}
\usepackage[T1]{fontenc}
\usepackage[ansinew]{inputenc}
\usepackage{graphics}
\usepackage{dcolumn}
\usepackage{bm}
\usepackage{amsthm}
\usepackage{amsmath}
\usepackage{amssymb}
\usepackage{diagbox}
\usepackage{hyperref}
\usepackage{url}
\usepackage{xcolor}

\begin{document}

\title{Understanding two-dimensional tractor magnets: theory and realizations}
 
\author{Michael P.\ Adams}\email[Electronic address: ]{michael.adams@uni.lu}
\affiliation{Department of Physics and Materials Science, University of Luxembourg, 162A~avenue de la Faiencerie, L-1511~Luxembourg, Grand Duchy of Luxembourg}
 

\begin{abstract}
We present a comparative investigation of two-dimensional tractor magnet configurations, analyzing both theoretical predictions and experimental results with a focus on the minimal tractor magnet. The minimal tractor magnet consists of a rigid assembly of one attracting magnet (attractor), two repelling magnets (repulsors), and a fourth magnet (follower) that is magnetically stabilized in a local energy minimum. The theoretical framework relies on magnetostatics and stability analysis of stationary equilibria. To calculate the magnetostatic force and energy, we use a multipole method. In a first approximation, we derive analytical results from the point dipole approximation. The point dipole analysis defines an upper bound criterion for the magnetic moment ratio and provides analytical expressions for stability bounds in relation to geometry parameters. Our experimental results are consistent with the predictions from the fourth-order multipole expansion. Beyond the minimal tractor magnet, we introduce a more advanced configuration that allows for a higher magnetic binding energy and follower capture at larger distances.

\end{abstract}

\date{\today}

\maketitle


\section{Introduction}

Magnetic stability is famously difficult to achieve.~\cite{Zangwill2012,berry1997flying,brandt1989levitation,geim1999magnet,kelvin1884reprint} Here we present the counterintuitive phenomenon of a tractor magnet: an assembly of three magnets fixed in place by a 3D mold that creates a stable trap for a fourth "follower" magnet. Achieving and refining stable two-dimensional magnet arrangements poses significant challenges, particularly in the presence of non-negligible friction forces. Moreover, the calculation of magnetic forces between real magnets often requires time-consuming computer simulations. In our study, we conducted both experimental and theoretical investigations into two-dimensional tractor magnets and their magnetic equilibria. The tractor magnet experiments presented may serve many uses, providing an engaging and cost-effective means for students to gain hands-on experience in computing stable equilibria. 

The paper is organized as follows: In Section~\ref{sec:Section2} we describe magnetostatic magnet-magnet interactions and the related stability analysis. In Section~\ref{sec:Section3} we discuss our findings on the stability behavior of the minimal tractor magnet configuration and present the results of our case study. In Section~\ref{sec:Section4} we provide the conclusions and an outlook. The supplementary material to this paper features a video file showing the different dynamic behaviors, the stl-files for 3D printing, MATLAB code, and a collection of selected tractor magnet designs with detailed information on the geometrical arrangement.~\cite{michael2024AJP}

\section{Magnetostatics and Stability}
\label{sec:Section2}

We consider a magnet with magnetization $\mathbf{M}_0$ (follower magnet) under influence of the magnetic field of a static assembly of $n$ magnets with magnetizations $\mathbf{M}_k$ (tractor magnet assembly). We assume uniform magnetizations of the type $\mathbf{M} = \pm B^{\mathrm{r}}\mathbf{\hat{z}}/\mu_0 $, where the remanent flux density is $B^{\mathrm{r}} = \mu_0 M^{\mathrm{r}}$. The total applied magnetic flux density $\mathbf{B}_{\mathrm{a}}$ related to a magnetizations $\mathbf{M}_k$ is described by
\begin{align}
\mathbf{B}_{\mathrm{a}}(\mathbf{r}) = 
\frac{\mu_0}{4\pi}\nabla_{\mathbf{r}}\times\nabla_{\mathbf{r}}\times \sum_{k=1}^{n}\int_{V_k}  \frac{\mathbf{M}_k(\mathbf{r}_k') \; d^3r_k'}{\|\mathbf{r}-\mathbf{r}_k'-\mathbf{r}_k\|}  ,
\label{eq:MagneticFluxDensityGeneral}
\end{align}
where $V_k$ is the $k$-th magnet volume, $\mathbf{r}$ is the field point, $\mathbf{r}_k$ is the center point of the $k$-th magnet, $\mathbf{r}_k'$ is the integration variable over the $k$-th magnet volume, $\mu_0$ is the vacuum permeability, and $\nabla_{\mathbf{r}}$ denotes the del-operator. The potential magnetostatic energy of the follower is described by
\begin{align}
    U_0(\mathbf{r}_0) = -  \int_{V_0} \mathbf{M}_0(\mathbf{r}'') \cdot \mathbf{B}_{\mathrm{a}}(\mathbf{r}_0 +\mathbf{r}'') d^3r'',
\label{eq:PotentialEnergyMangnet0}
\end{align}
where $\mathbf{r}_0$ is the geometrical center of the follower and $\mathbf{r}''$ is the integration variable with respect to the follower volume $V_0$. 
The magnetostatic force acting on the follower magnet is then given by the negative gradient of the potential energy
\begin{align}
\mathbf{F}_{0}(\mathbf{r}_0) = -\nabla_{\mathbf{r}_0} U_0(\mathbf{r}_0)
\label{eq:MagnetostaticForce_General}.
\end{align}
The satisfactory conditions for stability in a two-dimensional plane are
\begin{align}
    \left.\frac{\partial^2 U_0}{\partial x^2}\right|_{\mathbf{r}=\mathbf{r}_{\mathrm{eq}}}&> 0
    , &
    \left.\frac{\partial^2 U_0}{\partial y^2}\right|_{\mathbf{r}=\mathbf{r}_{\mathrm{eq}}}&> 0.
\end{align}
In our case the second order partial derivatives are assumed to be non-zero.
\begin{figure}[tb!]
\centering
\resizebox{1.0\columnwidth}{!}{\includegraphics{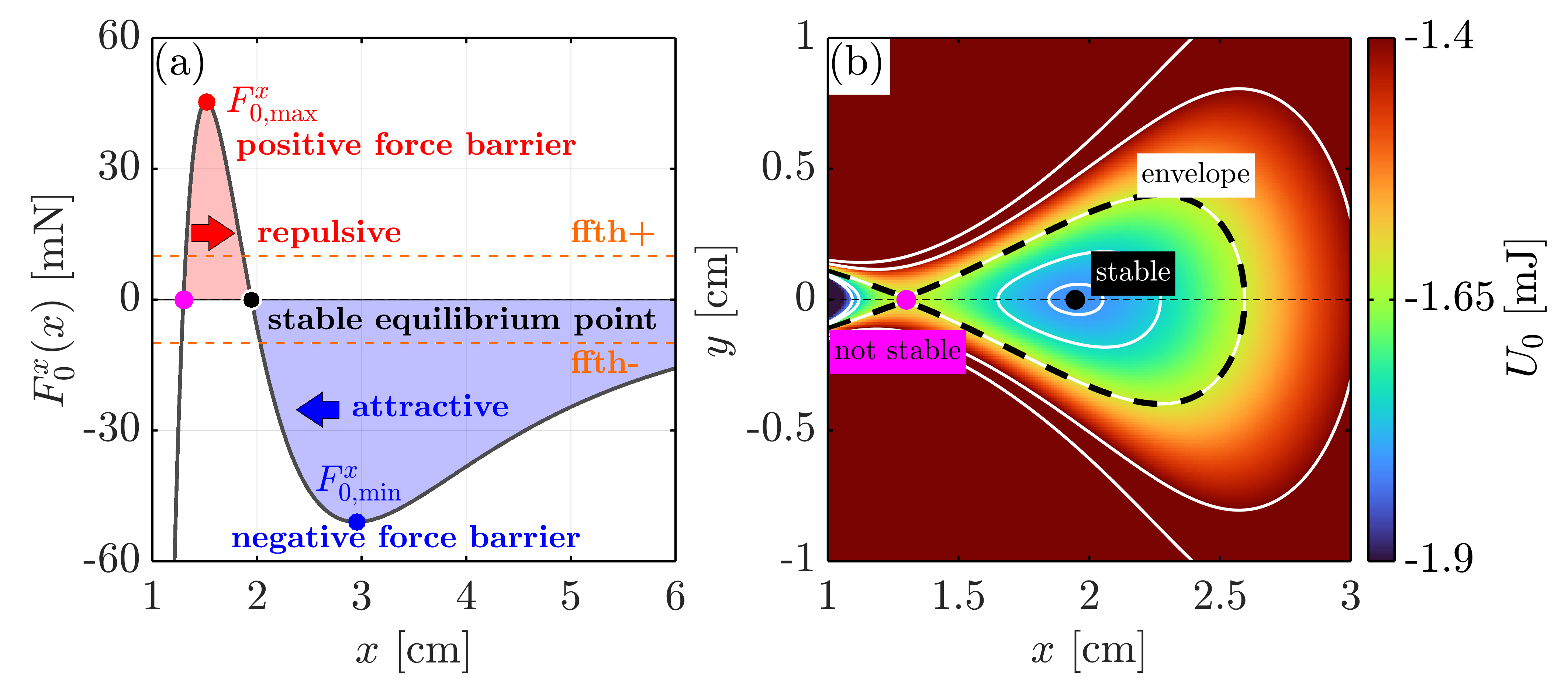}}
\caption{Example of a minimal tractor magnet with stationary stable equilibrium point [compare with Fig.~\ref{fig2}]. Panel~(a) shows the $x$-component of the magnetic force function where the black dot indicates the stationary stable equilibrium point and the magenta dot the unstable equilibrium point. The dashed orange lines show the static friction force threshold (ffth$\pm$). Panel~(b) displays the corresponding two-dimensional potential energy landscape. The black-white dashed line highlights the critical energy level crossing the unstable equilibrium point (saddle point) marked with the magenta dot.}
\label{fig1}
\end{figure}
In practice, the functionality of a tractor magnet device depends on friction forces that are not included in the magnetostatic energy, which is why we use the following recipe for the identification of a experimentally achievable equilibrium point: For a given magnet setup (i)~compute the $F_{0}^{x}(x,0)$ force function along the $x$-direction, (ii)~identify whether a sign-change of $F_{0}^{x}$ exists that indicates a zero point $x_{\mathrm{eq}}$ with negative slope, (iii)~refine the estimate of the zero point $x_{\mathrm{eq}}$ by using an iterative Newton algorithm, (iv)~check whether the extrema of the positive and negative force barrier of $F_{0}^{x}$ overcome a certain constant static friction force threshold, (v)~and repeat these steps for $F_{0}^{y}$. In this work we use this procedure to compute a stability map as a function of the geometry parameters of the minimal tractor magnet configuration and compare it to the experimental stability map obtained from a case study with real magnets. Figure~\ref{fig1} illustrates a possible $x$-$y$ energy landscape and related force function component $F_0^{x}(x)$ resulting from the magnetostatic potential energy described by Eq.~\eqref{eq:PotentialEnergyMangnet0}. We emphasize that the magnet configurations in our study are always assumed to be mirror symmetric relative to the $x$-axis like shown in Fig.~\ref{fig3}(a), implying that the possible energy minimum is always located on the $x$-axis. Fig.~\ref{fig1}(a) shows that the force function can be attractive or repulsive.

\section{Results and Discussion}
\label{sec:Section3}

\begin{figure}[tb!]
\centering
\resizebox{1.0\columnwidth}{!}{\includegraphics{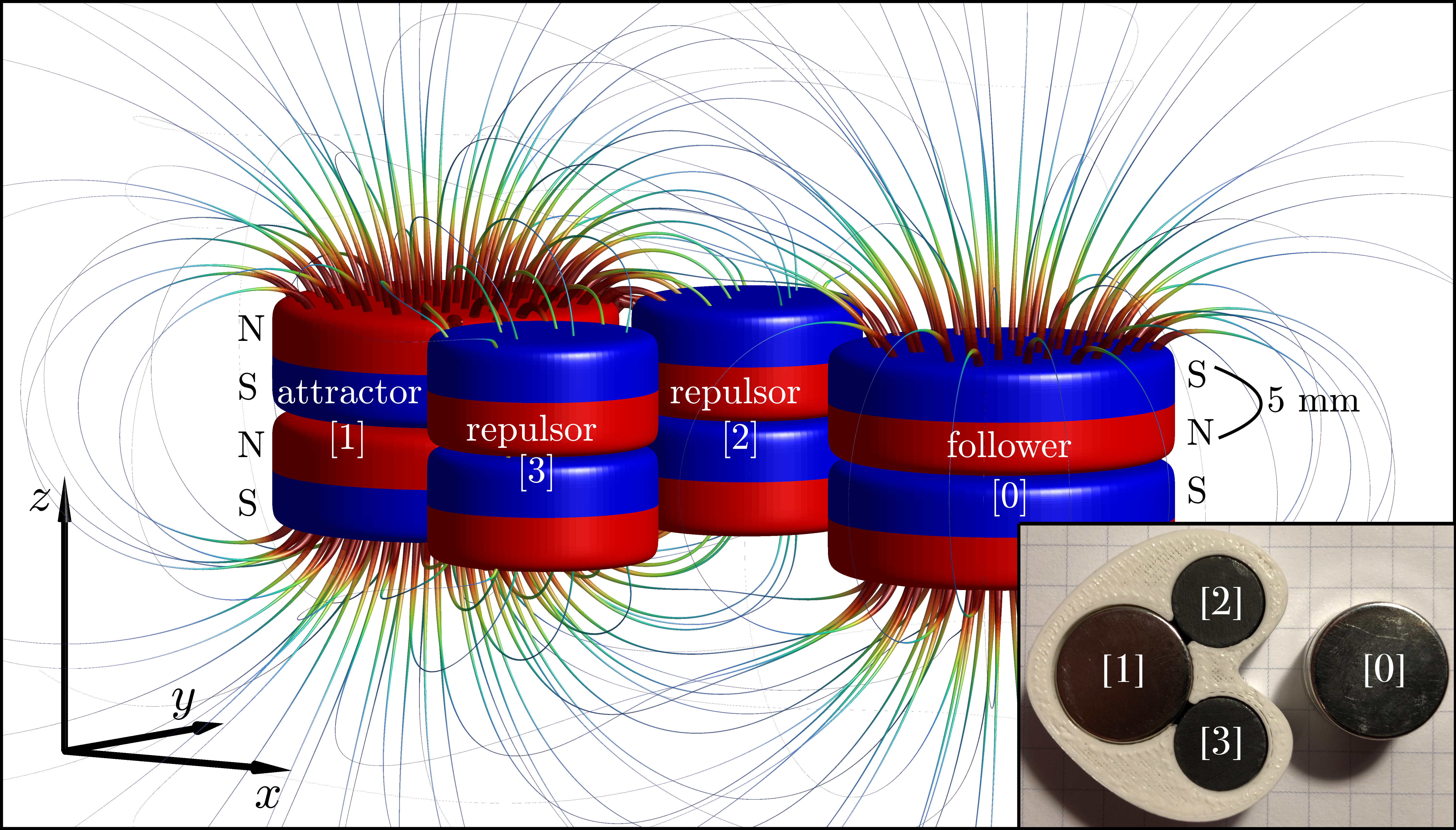}}
\caption{Minimal tractor magnet configuration. The follower magnet $[0]$  is attracted by the attractor magnet $[1]$ and repelled by the repulsor magnets $[2]$ and $[3]$. The follower and the attractor are Nd-Fe-B magnets and the repulsors are ferrite magnets.
}
\label{fig2}
\end{figure}

In our main study we consider the minimal tractor magnet configuration, which consists of four magnets: three magnets in the tractor magnet (one attractor magnet and two repulsor magnets) and one follower magnet. Our primary focus involves the investigation of the stationary stability behavior of this magnet arrangement. Figure~\ref{fig2} illustrates a minimal tractor magnet configuration accompanied by a corresponding 3D magnetic field line image. The inset depicts a realization of this system. The repulsors and the follower have the same orientation of the magnetization, while the attractor has the opposite orientation. In our configuration, both the attractor and the follower are Nd-Fe-B magnets, while the repulsors are ferrite magnets.

\begin{figure}[tb!]
\centering
\resizebox{1.0\columnwidth}{!}{\includegraphics{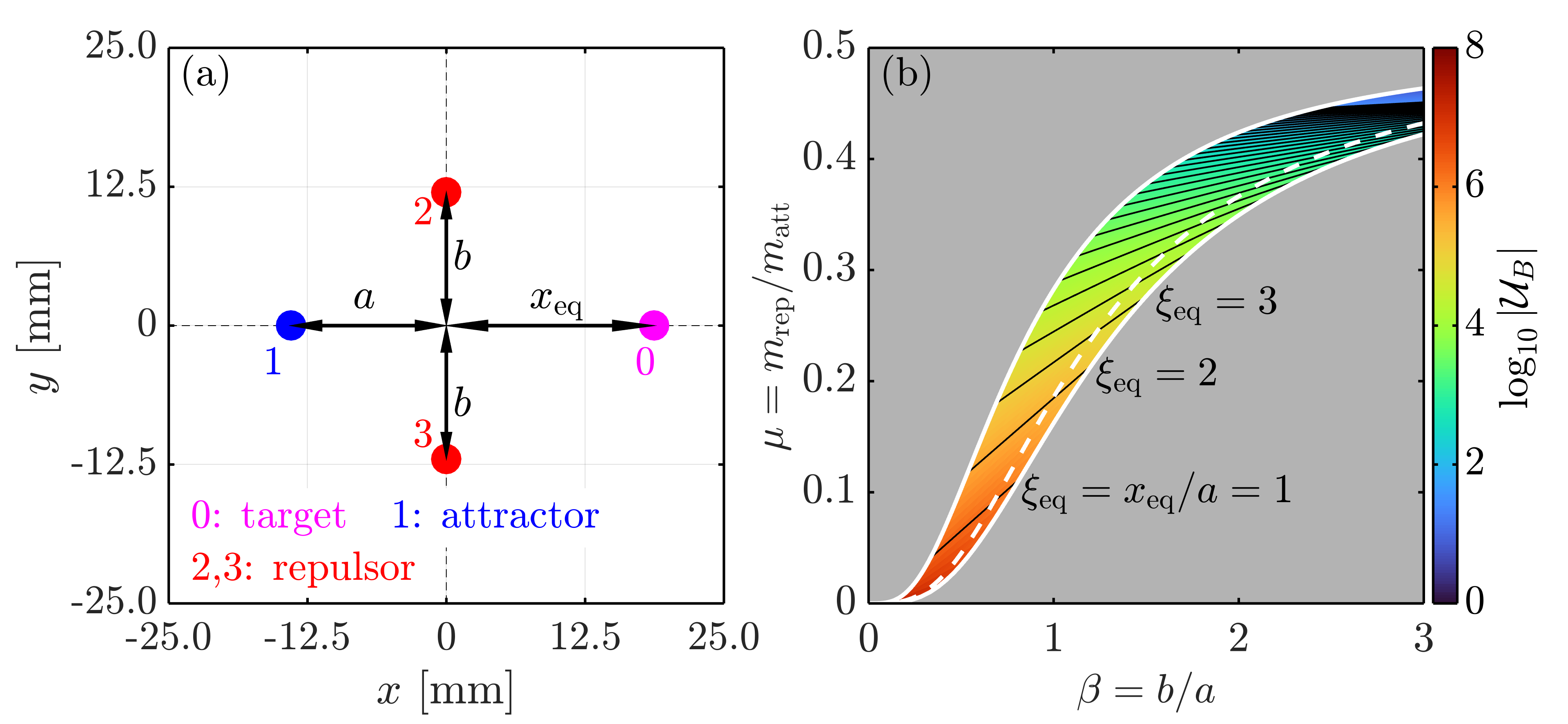}}
\caption{Result of the stability analysis under the point dipole approximation. Panel~(a) illustrates the configuration of four point dipoles. The magnetic moments of the dipoles are oriented parallel/antiparallel respectively to the $z$-direction (out of plane). The magnetic moments of the follower ($i=0$) and the attractor ($i=1$) dipole are antiparallel, while the magnetic moments of the repulsors ($i=2,3$) and the follower ($i=0$) are parallel. The magnitudes of the magnetic moments are defined as $m_{\mathrm{rep}} =|m_2| = |m_3|$ and $m_{\mathrm{att}} = |m_0| = |m_1|$. Panel~(b) displays the result of the stability analysis. The point dipole approximation predicts that the parameter configurations in the colored region are stable. The upper and lower bound of the stable region, indicated by the white solid lines, are computed by Eqs.~\eqref{eq:StabilityMapSurfaceParametrization} with $\lambda=0, \; \lambda=1$ respectively. Along the white dashed line the system is equally stable in $x$ and $y$-direction ($\partial^2U_0/\partial y^2 = \partial^2U_0/\partial x^2$).}
\label{fig3}
\end{figure}

The diameters are $d_{\mathrm{Nd}} = 15 \, \mathrm{mm}$ and $d_{\mathrm{Fe}} = 10 \, \mathrm{mm}$ and the heights are all $h = 2 \times 5 \, \mathrm{mm}$ (stack of two magnets). 
The remanent flux densities given by the manufacturer are $B_{\mathrm{Nd}}^{\mathrm{r}} = \mu_0 M_{\mathrm{Nd}}^{\mathrm{r}} = 1.3 \, \mathrm{T}$ for the Nd-Fe-B magnets and $B_{\mathrm{Fe}}^{\mathrm{r}} = \mu_0 M_{\mathrm{Fe}}^{\mathrm{r}}  = 0.39 \, \mathrm{T}$ for the ferrite magnets. These flux densities satisfy the upper bound criterion ($m = M V$) for the magnetic moment ratio 
\begin{align}
    \mu = \frac{m_{\mathrm{rep}}}{m_{\mathrm{att}}}=\frac{d_{\mathrm{Fe}}^2 B_{\mathrm{Fe}}^{\mathrm{r}}}{d_{\mathrm{Nd}}^2 B_{\mathrm{Nd}}^{\mathrm{r}}}
   = \frac{2}{15} < \frac{1}{2}.
    \label{eq:CriticalUpperBoundCriterion}
\end{align}
This criterion results from the fact that we have two repulsor magnets and one attractor magnet in our system. Equation~\eqref{eq:CriticalUpperBoundCriterion} can also be verified through the stability analysis of the point dipole approximation described in detail in Appendix~\ref{sec:PointDipoleModelApproximationAppendix}. 
Further statements resulting from the stability analysis of the point dipole approximation are taken from Fig.~\ref{fig3}. In Figure~\ref{fig3}(a) we introduce the geometry parameters $a$ (distance from the origin to the center of the attractor), $b$ (distance from the origin to the center of a repulsor) and $x_{\mathrm{eq}}$ (distance from the origin to the center of the follower). Figure~\ref{fig3}(b) displays the stability map in the plane of the magnetic moment ratio $\mu = m_{\mathrm{rep}}/m_{\mathrm{att}}$ and the geometrical ratio $\beta = b/a$ resulting from the point dipole approximation. The colormap represents the decadic logarithm of the dimensionless binding energy $\mathcal{U}_B$ (for details see Appendix~\ref{sec:PointDipoleModelApproximationAppendix}), where the black solid lines highlight stepwise integer levels of the parameter $\xi_{\mathrm{eq}}=x_{\mathrm{eq}}/a$. From Fig.~\ref{fig3}(b) we can deduce the following: (i)~in the gray area the tractor magnet configuration is predicted to be unstable, (ii)~the solid white lines highlight the transition from the stable to the unstable regions, (iii)~along the dashed white line the stability in $x$ and $y$-direction are equal in strength (linear stability, $\partial^2U_0/\partial x^2 = \partial^2U_0/\partial y^2$) and (iv)
upon increasing the geometrical ratio $\xi_{\mathrm{eq}}$ the binding energy $\mathcal{U}_B$ is decreasing. Thus, configurations on the white dashed line in Fig.~\ref{fig3}(b) in the left lower corner are of particular interest, because they have a symmetric stability and relatively high binding energies.

The findings from the point dipole approximation are very helpful for an initial understanding of the principles of a minimal tractor magnet configuration, but the precision of this approximate calculation is not sufficient for the fine-tuning of the geometry. To achieve higher accuracy, we use a Cartesian multipole expansion method to calculate the magnetostatic energy and force. We describe the details of this method in Appendix~\ref{sec:CartesianMultipoleExpansion}, where we show in Fig.~\ref{fig5} that the multipole expansion up to the fourth order converges to the six order expansion.

\begin{figure}[tb!]
\centering
\resizebox{1.0\columnwidth}{!}{\includegraphics{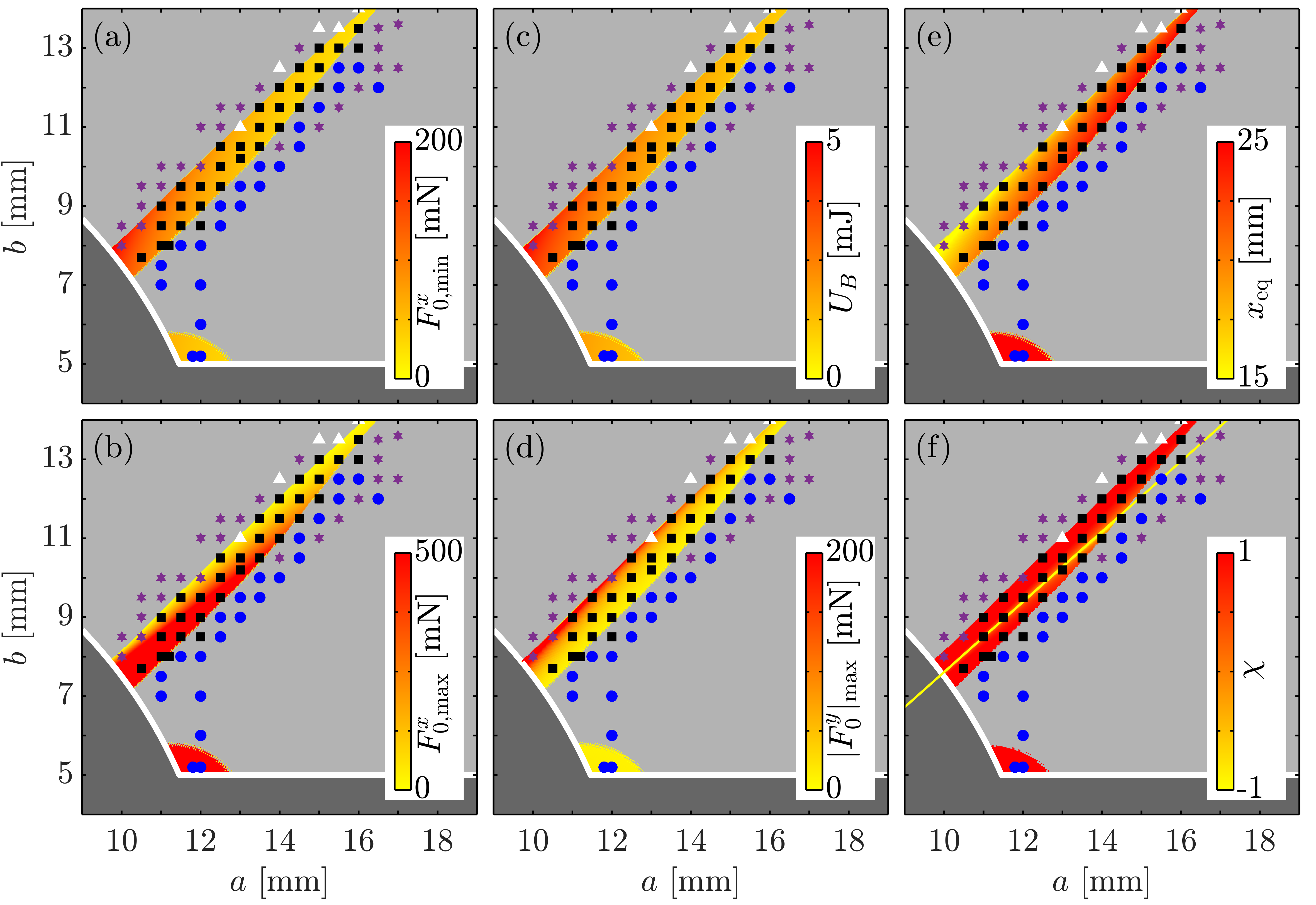}}
\caption{Results from our experimental case study of the minimal magnetic tractor configuration [see sketch in Fig.~\ref{fig3}(a)]. The dots represent the experimental data, and the underlying surfaces from~(a) to~(f) display the quantities from our theoretical predictions. Configurations $(a,b)$ in the light gray region are unstable, while the dark gray region is forbidden due to the geometrical limits. The static friction force threshold is estimated to be $F_{\mathrm{fric}}=10 \, \mathrm{mN}$. The comparison of the black squares (corresponding to stable tractor magnet configurations) with the predicted stability region (colored surface in the shape of an elongated shark fin) shows a good agreement between theory and experiment. 
}
\label{fig4}
\end{figure}

Figure~\ref{fig4} shows the results of our experimental case study alongside the theoretical predictions derived from the fourth-order multipole expansion. In this case study we produced 82 samples with a 3D printer (see example in the inset of Fig.~\ref{fig2}), with variation of the geometry parameters $a$ and $b$. We then checked manually whether a certain tuple $(a,b)$ is sufficient for the requirements of a tractor magnet, and made entries to Fig.~\ref{fig4}, where the markers represent the experimental data. The markers in Fig.~\ref{fig4} are understood as follows: (i)~the black squares indicate functioning tractor magnet configurations, (ii)~the blue dots indicate configurations where the force barrier in $y$-direction is not sufficient, (iii)~the white triangles indicate configurations where the force barrier in $x$-direction is not sufficient, and (iv)~the violet stars indicate configurations showing no local energy minima at all. The underlying colormaps in Figs.~\ref{fig4}(a)$-$(f) show different quantities from our theoretical calculations: (a)~displays the absolute value of the minimum of the negative force barrier of the $F_{0}^{x}$-force component [compare to Fig.~\ref{fig1}(a)], (b)~shows the maximum of the positive force barrier of the $F_{0}^{x}$-force component [compare to Fig.~\ref{fig1}(a)], (c)~displays the absolute value of the magnetic binding energy $U_B=|U_0(x_{\mathrm{eq}},0)|$, (d)~displays the maximum of the $F_{0}^{y}$-force component, (e)~displays the static equilibrium position $x_{\mathrm{eq}}$ of the follower magnet [compare Fig.~\ref{fig3}(a)], and (f)~displays the quantitiy $\chi = \log_{10}\left| \partial^2U_0/\partial y^2  -  \partial^2U_0/\partial x^2 \right|$ at the point $(x,y) = (x_{\mathrm{eq}},0)$, which is a measure for the symmetry of the stability in $x$ and $y$-direction (in the symmetric case the limit $\chi \rightarrow - \infty$ is achieved). As in Fig.~\ref{fig3}, the light gray background in Fig.~\ref{fig4} indicates the predicted unstable region that follows the recipe described in Sec.~\ref{sec:Section2} with a static friction force threshold of $F_{\mathrm{fric}} = 10 \; \mathrm{mN}$. The white solid line in Fig.~\ref{fig4} represents the geometrical boundaries, resulting from the fact that the magnets must not overlap:
\begin{align}
    \sqrt{a^2 + b^2} &> \frac{d_{\mathrm{Nd}} + d_{\mathrm{Fe}}}{2} = 12.5 \; \mathrm{mm} , & b &> d_{\mathrm{Fe}}/2 = 5 \;  \mathrm{mm}.
\end{align}
Due to these boundaries, the dark gray region is forbidden. As seen in Fig.~\ref{fig4}, the main stable region predicted from theory has the shape of an elongated shark fin with a width of approximately $1$ mm, and apart from a few outliers, our prediction is in very good agreement with the experimental data (black dots are stable). The pocket located in the region $5<b/\mathrm{mm}<6$ is predicted to be stable, but in the experiment we observe that it is not stable. It is not yet clear to us what the origin of this artificial pocket is. 

For the design of a ``good'' tractor magnet (where good means symmetric and having a strong binding energy), Figs.~\ref{fig4}(c), (e) and (f) are of highest interest. In Fig.~\ref{fig4}(f) we surprisingly find from our theoretical predictions a ``line of symmetry'' in the $(a,b)$-stability map that is approximately described by the linear function [solid yellow line in Fig.~\ref{fig4}(f)]:
\begin{align}
b_{\mathrm{mid}}(a) \approx 0.89 a - 1.29 \; \mathrm{mm}   , \quad 10 < a/\mathrm{mm} < 14.5.
\label{eq:LineOfStability}
\end{align}
Along this path, the difference of the second-order derivatives of the potential magnetic energy  vanishes [$\partial^2U_0/\partial y^2 = \partial^2U_0/\partial x^2$], indicating symmetric stability. In Fig.~\ref{fig4}(c) we see that the binding energy $U_B$ is relatively strong for small $a\sim 10 \, \mathrm{mm}$, making this range interesting for practical applications. By contrast [see Fig.~\ref{fig4}(e)] the equilibrium distance $x_{\mathrm{eq}}$ is decreasing with a decreasing geometry parameter $a$. This makes sense, since the potential energy decreases with increasing distance between the magnets.

\section{Conclusions and Outlook}
\label{sec:Section4}

We investigated 2D magnet configurations with a stable stationary equilibrium point. Based on the theory of magnetostatics we have derived general statements on the working principle and we have provided practically useful information for the fine-tuning of the minimal tractor magnet configuration: (i)~we made a qualitative statement on magnet stacking and gave an explanation of the consequences for the binding energy based on the pole-avoidance principle, (ii)~we have introduced an upper bound criterion for the magnetic moment ratio ($\mu<1/2$), (iii)~based on a point dipole approximation, we have analytically derived a stability map of the problem that allows one to identify the practically important regime of the magnetic moment ratio $\mu=m_{\mathrm{rep}}/m_{\mathrm{att}}$ and the geometry parameter ratio $\beta=b/a$, (iv)~we have demonstrated that the theoretical predictions (based on a fourth-order multipole expansion) are in qualitative agreement with the experimental stability map (obtained with 82 samples), (v)~our fourth-order multipole calculation results into a linear path in the $(a,b)$-stability map, along which the linear stability is equal in strength with respect to the $x$- and $y$-directions, and (vi)~we have introduced an advanced tractor magnet composed of two minimal tractor magnets and discussed the possibility of much higher binding energies in such arrangements. Our results establish a connection between the magnet geometry, characterized by size parameters such as the diameter $d$ and height $h$, the spatial arrangement of the magnets (geometrical parameters $a$ and $b$), and the magnetization magnitude $M^r$, elucidating their collective influence on stability properties. In the supplementary Material we provide a video where we show different tractor magnets and showcase their stability behavior on a qualitative level.

\acknowledgments{I thank the National Research Fund of Luxembourg for financial support (AFR Grant No.~15639149). I thank Prof.\ Andreas Michels, Prof.\ Hamid Kachkachi, Dr. Venus Rai and Dr.\ Ivan Titov for fruitful discussions and the critical reading of the manuscript.}


%

\appendix

\section{Stability analysis of the point dipole model}\label{sec:PointDipoleModelApproximationAppendix}

In the first approximation, a point dipole model provides useful insights into the stability behavior of magnetic tractors. Figure~\ref{fig3}(a) shows the arrangement of four magnetic point dipoles, where the dipole with index $i=0$ represents the follower magnet and the assembly dipoles with indices $i=1,2,3$ represent the tractor magnet. The tractor magnet consists of an attractor magnet ($i=1$) and two repulsor magnets ($i=2,3$). The polarization of the dipoles is parallel or antiparallel to the $z$-direction (out of plane). The corresponding positions $\mathbf{r}_i$ and magnetic moments $\mathbf{m}_i$ are:
\begin{align}
\mathbf{r}_0 &= x\mathbf{\hat{x}} + y\mathbf{\hat{y}}, & \mathbf{m}_0 &= m_{\mathrm{att}} \mathbf{\hat{z}},
\\
\mathbf{r}_1 &= -a \mathbf{\hat{x}}, & \mathbf{m}_1 &= -m_{\mathrm{att}} \mathbf{\hat{z}},
\\
\mathbf{r}_2 &= b \mathbf{\hat{y}} ,& \mathbf{m}_2 &= m_{\mathrm{rep}} \mathbf{\hat{z}},
\\
\mathbf{r}_3 &=-b \mathbf{\hat{y}}, & \mathbf{m}_3 &= m_{\mathrm{rep}} \mathbf{\hat{z}},
\end{align}
where $a,b>0$ are the position parameters of the tractor magnet. The sign convention follows the interpretation of the attractor and repulsor magnets, thus, the repulsor magnets ($i=2,3$) are parallel and the attractor magnet ($i=1$) is antiparallel in its magnetization relative to the follower magnet ($i=0$). In the case of a magnet with uniform magnetization parallel/antiparallel to the $z$-direction and a cylindrical volume $V_c = \pi R^2 h$, the magnetic moment is written as:
\begin{align}
    \mathbf{m} = \int_{V} \mathbf{M}(\mathbf{r}) \; d^3r  =  \pm\frac{\pi R^2 h B^{\mathrm{r}}}{\mu_0}\mathbf{\hat{z}},
\end{align}
where $R$ is the radius, $h$ denotes the height of the magnet, $M^{\mathrm{r}} = B^{\mathrm{r}}/\mu_0$ is the remanent magnetization. The magnetic flux density at the position $\mathbf{r}_0$ of the follower magnet ($i=0$) is:
\begin{align}
    \mathbf{B}_0(x,y) &= \frac{\mu_0}{4\pi} \sum_{n=1}^{3}\left[ 3 \frac{(\mathbf{R}_n\cdot\mathbf{m}_n) \mathbf{R}_n}{R_n^5} - \frac{\mathbf{m}_n}{R_n^3} \right]  ,
\end{align}
where $\mathbf{R}_n = \mathbf{r}_0 - \mathbf{r}_n$ is the difference vector between the field point and the source point, and $R_n = \sqrt{(R_n^{x})^2 + (R_n^{y})^2 + (R_n^{z})^2}$ is the magnitude of $\mathbf{R}_n$. The potential energy then reads:
\begin{align}
    U_0(x,y) &= 
    - \mathbf{m}_0 \cdot \mathbf{B}_0(x,y)
    \\
    &=
    \frac{\mu_0 m_{\mathrm{att}} m_{\mathrm{rep}}}{4\pi} 
    \left[\frac{1}{(x^2 + (y-b)^2)^{3/2}}  + \frac{1}{(x^2 + (y+b)^2)^{3/2}} \right] \nonumber
    \\
    &-\frac{\mu_0 m_{\mathrm{att}}^2}{4\pi} \frac{1}{((x+a)^2 + y^2)^{3/2}}.
\label{eq:PotentialEnergy_FourPointDipoleModel}
\end{align}
By analyzing the first-order stability conditions we find that the following relation between the magnetic moment ratio $\mu = m_{\mathrm{rep}}/m_{\mathrm{att}}$ and the geometry parameters must be satisfied:
\begin{align}
    \frac{\partial U_0}{\partial x}(x_{\mathrm{eq}}, 0) &= 0
    \quad \rightarrow \quad \mu = \frac{m_{\mathrm{rep}}}{m_{\mathrm{att}}} = \frac{(x_{\mathrm{eq}}^2 + b^2)^{5/2}}{2 x_{\mathrm{eq}}(x_{\mathrm{eq}} + a)^4}
    \\
    \frac{\partial U_0}{\partial y}(x_{\mathrm{eq}}, 0) &= 0 \quad \rightarrow \quad \mathrm{satisfied \; due \; to \; symmetry}
\end{align}
The second-order stability conditions provide lower and upper bound inequalities for the geometry parameters:
\begin{align}
   \frac{\partial^2 U_0}{\partial x^2}(x_{\mathrm{eq}}, 0)  &= C_0[4 a x_{\mathrm{eq}}^2 - b^2 (a + 5 x_{\mathrm{eq}})] >0 \quad \rightarrow \quad b < b_{\mathrm{up}} = \sqrt{\frac{4 a x_{\mathrm{eq}}^2}{5x_{\mathrm{eq}} + a}}
    \\
    \frac{\partial^2 U_0}{\partial y^2}(x_{\mathrm{eq}}, 0) &= C_0[b^2 (5 x_{\mathrm{eq}} + 4 a) - a x_{\mathrm{eq}}^2] > 0 \quad \rightarrow \quad b > b_{\mathrm{lo}} = \sqrt{\frac{a x_{\mathrm{eq}}^2}{5 x_{\mathrm{eq}} + 4 a}}
\\
    \frac{\partial^2 U_0}{\partial x^2}(x_{\mathrm{eq}}, 0) &= \frac{\partial^2 U_0}{\partial y^2}(x_{\mathrm{eq}}, 0) > 0   \quad \rightarrow \quad  b_{\mathrm{mid}} = \sqrt{\frac{a x_{\mathrm{eq}}^2}{2x_{\mathrm{eq}} + a}}
    \\
    C_0 &= \frac{3\mu_0 m_{\mathrm{att}}^2}{4\pi x_{\mathrm{eq}} (x_{\mathrm{eq}} + a)^5 (x_{\mathrm{eq}}^2 + b^2)}
\end{align}
The parameter $b_{\mathrm{mid}}$ gives the condition that the slope of the $x$- and $y$-force components are equal and thus describes the middle path between the lower and upper bound. Using the reduced parameters $\xi_{\mathrm{eq}} = x_{\mathrm{eq}}/a$, $\beta = b/a$, and $\mu = m_{\mathrm{rep}}/m_{\mathrm{att}}$ we can express the lower bound $\beta_{\mathrm{lo}}$, the upper bound $\beta_{\mathrm{up}}$, the middle path $\beta_{\mathrm{mid}}$, and the magnetic moment ratio $\mu$ as:
\begin{align}
    \beta_{\mathrm{lo}} &= \frac{\xi_{\mathrm{eq}}}{\sqrt{5\xi_{\mathrm{eq}} + 4}},
    &
    \beta_\mathrm{up} &= \frac{2\xi_{\mathrm{eq}}}{\sqrt{5\xi_{\mathrm{eq}} + 1}},
    \label{eq:DipoleApprox_ScaledStabilitySolution1}
    \\
    \beta_{\mathrm{mid}} &= \frac{\xi_{\mathrm{eq}}}{\sqrt{2\xi_{\mathrm{eq}} + 1}},
    &
    \mu &= \frac{(\xi_{\mathrm{eq}}^2 + \beta^2)^{5/2}}{2\xi_{\mathrm{eq}} (\xi_{\mathrm{eq}}+1)^4}.
\label{eq:DipoleApprox_ScaledStabilitySolution2}
\end{align}
By inversion of the relations  \eqref{eq:DipoleApprox_ScaledStabilitySolution1} and  \eqref{eq:DipoleApprox_ScaledStabilitySolution2} we find the lower and upper bound for the magnetic moment ratio $\mu$ depending on the scaled parameter $\xi_{\mathrm{eq}}$: 
\begin{align}
\mu_{\mathrm{lo}} &= \frac{\xi_{\mathrm{eq}}^4}{2 (\xi_{\mathrm{eq}} + 1)^4} \left[\frac{5\xi_{\mathrm{eq}} + 5}{5\xi_{\mathrm{eq}} + 4} \right]^{5/2}
\label{eq:UpperBoundMagnMomRatio},
\\
\mu_{\mathrm{up}} &= \frac{\xi_{\mathrm{eq}}^4}{2 (\xi_{\mathrm{eq}} + 1)^4} \left[\frac{5\xi_{\mathrm{eq}} + 5}{5\xi_{\mathrm{eq}} + 1} \right]^{5/2}
\label{eq:LowerBoundMagnMomRatio},
\\
\mu_{\mathrm{mid}} &= \frac{\xi_{\mathrm{eq}}^4}{2 (\xi_{\mathrm{eq}} + 1)^4} \left[\frac{2\xi_{\mathrm{eq}} + 2}{2\xi_{\mathrm{eq}} + 1} \right]^{5/2}
\label{eq:MidBoundMagnMomRatio}.
\end{align}
Computing the limits of Eqs.~\eqref{eq:UpperBoundMagnMomRatio} and \eqref{eq:LowerBoundMagnMomRatio} we find:
\begin{align}
\lim_{\xi_{\mathrm{eq}}\rightarrow 0} \mu_{\mathrm{lo}} &= 0,  &
\lim_{\xi_{\mathrm{eq}}\rightarrow 0} \mu_{\mathrm{up}} &= 0,
\\
\lim_{\xi_{\mathrm{eq}}\rightarrow \infty} \mu_{\mathrm{lo}} &= \frac{1}{2},  &
\lim_{\xi_{\mathrm{eq}}\rightarrow \infty} \mu_{\mathrm{up}} &= \frac{1}{2},
\end{align}
leading to the following critical bounds for the magnetic moment ratio:
\begin{align}
     0 < \mu  < \frac{1}{2}.
\end{align}
This result indicates that the magnetic dipole moment of the repulsor magnets should be less than half of the magnetic dipole moment of the attractor magnet. Further, these limits suggest that if we would like to have a large holding distance ratio $\xi_{\mathrm{eq}} = x_{\mathrm{eq}}/a$, the magnetic moment ratio $\mu$ must be close to $1/2$. In contrast the reduced binding energy:
\begin{align}
    \mathcal{U}_B(\xi_{\mathrm{eq}}, \beta) = 
    \frac{4\pi a^3 }{\mu_0 m_{\mathrm{att}}^2} U_0(x_{\mathrm{eq}}, 0) 
    &= \frac{\xi_{\mathrm{eq}}^2 + \beta^2}{\xi_{\mathrm{eq}}(\xi_{\mathrm{eq}}+1)^4}
    - \frac{1}{(\xi_{\mathrm{eq}} + 1)^3},
\end{align}
vanishes in the limit $\xi_{\mathrm{eq}} \rightarrow \infty$, showing the competition between holding-distance and binding energy. Combining the results for the upper bounds $\beta_{\mathrm{up}}, \mu_{\mathrm{up}}$ and lower bounds $\beta_{\mathrm{lo}}, \mu_{\mathrm{lo}}$, we can describe the region of stability $\mathbf{S}$ in the $(\beta,\mu)$-map by introducing the dimensionless parameter $0\le \lambda \le 1$ by the parametrization
\begin{align}
    \mathbf{S}(\xi_{\mathrm{eq}}, \lambda)
    =
    \begin{bmatrix}
    \beta(\xi_{\mathrm{eq}}, \lambda)
    \\
    \mu(\xi_{\mathrm{eq}} , \lambda)
    \end{bmatrix}
    = 
    \left\{
    \begin{bmatrix}
    \beta_{\mathrm{up}}(\xi_{\mathrm{eq}})
    \\
    \xi_{\mathrm{up}}(\xi_{\mathrm{eq}})
    \end{bmatrix}
    -
    \begin{bmatrix}
    \beta_{\mathrm{lo}}(\xi_{\mathrm{eq}})
    \\
    \xi_{\mathrm{lo}}(\xi_{\mathrm{eq}})
    \end{bmatrix}\right\}
    \lambda 
    +
    \begin{bmatrix}
    \beta_{\mathrm{lo}}(\xi_{\mathrm{eq}})
    \\
    \xi_{\mathrm{lo}}(\xi_{\mathrm{eq}})
    \end{bmatrix}.
\label{eq:StabilityMapSurfaceParametrization}
\end{align}
Here, $\lambda$ is a free parameter that spans the stability surface between the lower and upper bounds (see Fig.~\ref{fig3}). This mapping is essential for making statements on the stability of the tractor magnet configuration, since it reflects the stability behavior directly related to the magnetic-moment ratio $\mu=m_{\mathrm{rep}}/m_{\mathrm{att}}$ and the geometrical ratio $\beta = b/a$.

\section{Cartesian multipole expansion method}\label{sec:CartesianMultipoleExpansion}

To formulate the potential energy of the magnetostatic interaction in terms of the multipole expansion, we use the magnetic Hertz vector $\boldsymbol{\Pi}_k(\mathbf{r})$ (dipole vector potential) for the $k$-th magnet with magnetization $\mathbf{M}_k(\mathbf{r})$ in the convolution integral formulation~\cite{nisbet1955hertzian,essex1977hertz}
\begin{align}
    \boldsymbol{\Pi}_k(\mathbf{r}) &= \frac{1}{4\pi} \int_{V_k} \frac{\mathbf{M}_k(\mathbf{r}')}{\|\mathbf{r}-\mathbf{r}'\|} d^3r'. \label{eq:MagneticHertzVector}
\end{align}
The inverse relation between the magnetostatic Hertz vector and the magnetization vector field is described by the Poisson equation $\nabla^2\boldsymbol{\Pi}_k(\mathbf{r}) = - \mathbf{M}_k(\mathbf{r})$. The Cartesian multipole expansion of Eq.~\eqref{eq:MagneticHertzVector} is then written as~\cite{Jackson1999electrodynamics}
\begin{align}
     \boldsymbol{\Pi}_k(\mathbf{r})= \frac{1}{4\pi} \sum_{\boldsymbol{\nu}=\mathbf{0}}^{\boldsymbol{\infty}} \frac{(-1)^{|\boldsymbol{\nu}|} \boldsymbol{\mathcal{M}}_{k,\boldsymbol{\nu}}}{\boldsymbol{\nu}!} \left.\frac{\partial^{|\boldsymbol{\nu}|} (\|\mathbf{x}\|^{-1})}{\partial \mathbf{x}^{\boldsymbol{\nu}}}\right|_{\mathbf{x} = \mathbf{r}-\mathbf{r}_k},
\end{align}
where $\boldsymbol{\nu}=\{\nu_x, \nu_y, \nu_z\}$ is a multi-index, $|\boldsymbol{\nu}| = \nu_x + \nu_y + \nu_z$ denotes the absolute value of the multi-index, $\boldsymbol{\mathcal{M}}_{k,\boldsymbol{\nu}}$ are the mutipole moments of the magnetization $\mathbf{M}_k$, and $\mathbf{r}_k$ represents the position of the center of the $k$-th magnet. In the case of cylindrical magnets with radius $R_k$ and height $h$ with uniform magnetization vector field $\mathbf{M}_k = p_k B_k^{\mathrm{r}}/\mu_0 \mathbf{\hat{z}}, \; p_k\in\{-1,1\}$, the multipole moments read
\begin{align}
\boldsymbol{\mathcal{M}}_{k,\boldsymbol{\nu}} &= \int_{V_k} \mathbf{M}_k(\mathbf{r}) \mathbf{r}^{\boldsymbol{\nu}} \; d^3r
= 
p_k\frac{B_k^{\mathrm{r}}}{\mu_0} I_{k,\nu_x, \nu_y}^{\rho} I_{k,\nu_x, \nu_y}^{\phi} I_{k,\nu_z}^{z},
\end{align}
with the integrals
\begin{align}
    I_{k,\nu_x, \nu_y}^{\rho} &= \int_{0}^{R_k}  \rho^{\nu_x + \nu_y + 1} d\rho = \frac{R_k^{\nu_x+\nu_y+2}}{\nu_x + \nu_y + 2},
    \\
    I_{k,\nu_x,\nu_y}^{\varphi} &= \int_{0}^{2\pi}\cos^{\nu_x}\varphi \sin^{\nu_y}\varphi d\varphi = 2 \; \mathrm{B}\left( \frac{\nu_x+1}{2}, \frac{\nu_y+1}{2}\right) \kappa_{\nu_x} \kappa_{\nu_y},
    \\
    I_{k,\nu_z}^{z} &= \int_{-h/2}^{h/2} z^{\nu_z} dz = \frac{h^{\nu_z+1}\kappa_{\nu_z}}{2^{\nu_z}(\nu_z+1)},
    \\
    \kappa_n  &= \frac{1 + (-1)^n}{2},
\end{align}
where $\mathrm{B}(\cdot, \cdot)$ denotes the Euler beta function. The total magnetic flux density is then derived through the double-curl operation on the sum over the Hertz vectors associated with the related magnets,~\cite{nisbet1955hertzian,essex1977hertz}
\begin{align}
    \mathbf{B}(\mathbf{r}) &= \mu_0 \nabla\times \nabla \times \sum_{k=1}^{n}\boldsymbol{\Pi}_k(\mathbf{r}).
\end{align}
\begin{figure}[tb!]
\centering
\resizebox{0.50\columnwidth}{!}{\includegraphics{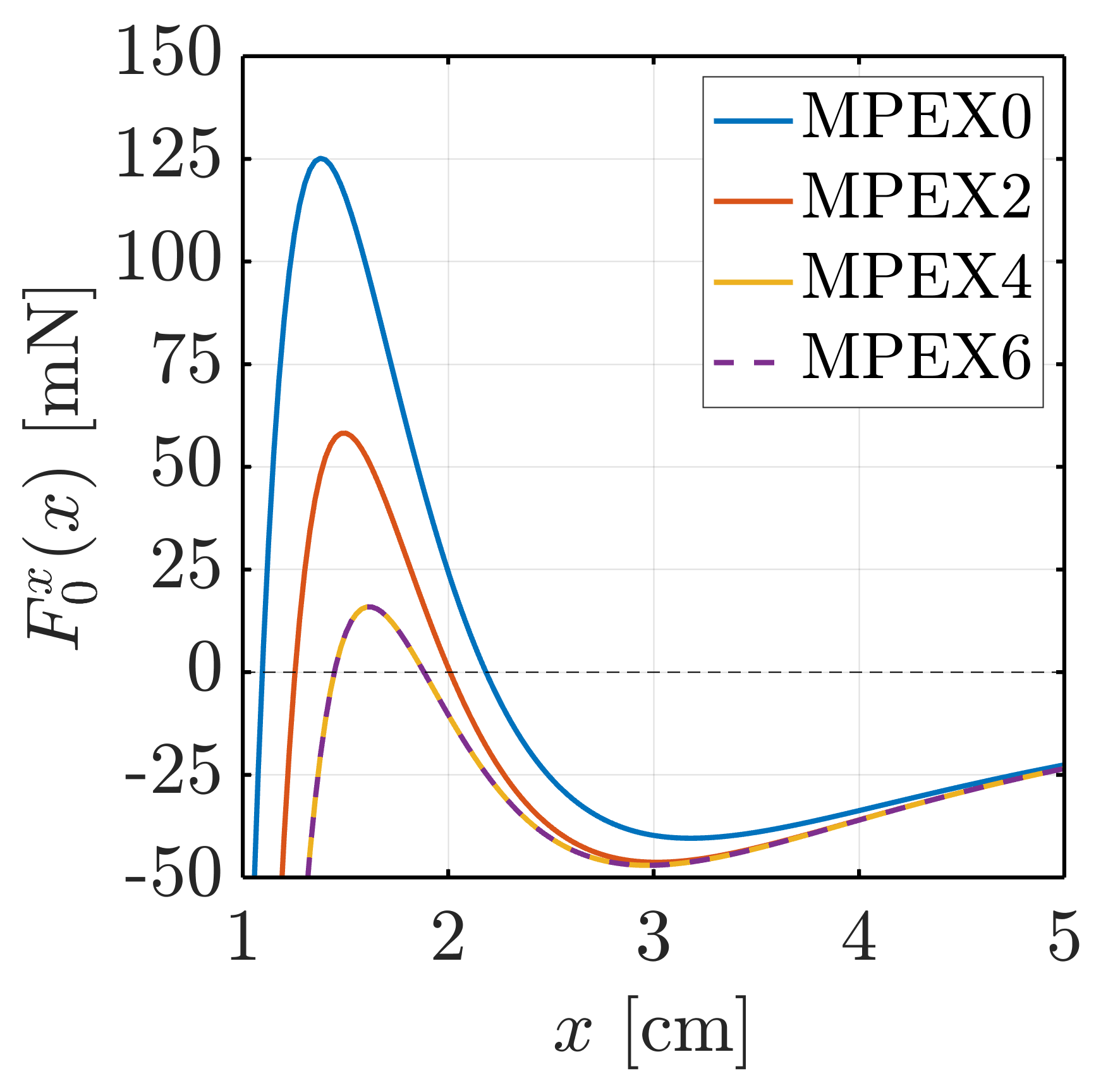}}
\caption{Convergence of the $F_{0}^{x}(x)$ force function computed with the multipole expansion method. MPEX0 illustratets the result from the four dipole model with potential energy \eqref{eq:PotentialEnergy_FourPointDipoleModel}. The graphs MPEX2,4 and 6 illustrate the results from the second to the six order multipole-expansion method described in Appendix~\ref{sec:CartesianMultipoleExpansion}. Here we used the parameters $a = 14$ mm, $b = 12$ mm, $R_0 = 7.5$ mm, $R_2 = 5$ mm, $B_0^{\mathrm{r}} = 1300$ mT and $B_2^{\mathrm{r}} = 390$ mT.}
\label{fig5}
\end{figure}
In the case of the tractor magnet with minimum number of magnets $n=3$, the center positions are
\begin{align}
\mathbf{r}_1 &= \{-a, 0, 0\},
\\
\mathbf{r}_2 &= \{0, b, 0\},
\\
\mathbf{r}_3 &= \{0, -b, 0 \}.
\end{align}
The potential energy of the follower magnet ($k=0$), with magnetization $\mathbf{M}_0$ and position $\mathbf{r}= \{x,y,0\}$, is described by the correlation integral:
\begin{align}
    U_0(\mathbf{r})  &=- \int_{V_0} \mathbf{M}_0(\mathbf{r}'') \cdot \mathbf{B}(\mathbf{r} + \mathbf{r}'') \; d^3 r'',
\end{align}
and finally the multipole expansion of this correlation integral reads
\begin{align}
    U_0(\mathbf{r})  &= - \sum_{\boldsymbol{\mu}=\mathbf{0}}^{\boldsymbol{\infty}} \frac{\boldsymbol{\mathcal{M}}_{0,\boldsymbol{\mu}}}{\boldsymbol{\mu}!} \cdot\frac{\partial^{|\boldsymbol{\mu}|} \mathbf{B}(\mathbf{r})}{\partial \mathbf{r}^{\boldsymbol{\mu}}}.
\end{align}

In our calculation, we first compute the potential energy analytically up to a certain order and then evaluate the resulting expression numerically. Figure~\ref{fig5} shows the convergence when increasing the order of the multipole expansion. The result MPEX0 ($\underline{\mathrm{m}}$ulti$\underline{\mathrm{p}}$ole $\underline{\mathrm{ex}}$pansion) reflects the point dipole approximation described by the potential energy in Eq.~\eqref{eq:PotentialEnergy_FourPointDipoleModel}. The higher-order multipole expansions (MPEX2, 4, 6) are computed following the equations described in this Appendix. At far distances, the graphs converge asymptotically, while for short distances, especially around the positive force barrier, the disagreement of MPEX0 and MPEX2 compared to MPEX4 is quite large. The results of MPEX4 and MPEX6 coincide. This demonstrates that the fourth-order multipole approximation provides a very good approximation. In the supplementary material the stability map~\ref{fig4} is shown as a result of the MPEX0, MPEX2 and MPEX4 method. These figures show that the MPEX4 method gives a prediction that is the closer to the experimental data.

\end{document}